\documentclass[doublecol]{epl2}

\usepackage{graphicx}
\usepackage{dcolumn}
\usepackage{bm}
\usepackage{color}
\usepackage{amsmath}
\usepackage{amssymb}

\title{Aging phenomena in the two-dimensional complex Ginzburg--Landau equation}
\shorttitle{Aging phenomena in the two-dimensional complex Ginzburg--Landau equation} 

\author{Weigang Liu \and Uwe C. T\"auber}
\shortauthor{W. Liu \and U. C. T\"auber}

\institute{Department of Physics \& Center for Soft Matter and Biological Physics (MC 0435),
Robeson Hall, 850 West Campus Drive, Virginia Tech, Blacksburg, VA 24061, USA}

\pacs{05.40.-a}{Fluctuation phenomena, random processes, noise, and Brownian motion}
\pacs{89.75.Da}{Systems obeying scaling laws}
\pacs{64.60.-i}{General studies of phase transitions}

\abstract{The complex Ginzburg--Landau equation with additive noise is a stochastic partial differential equation that describes a remarkably wide range of physical systems which include coupled non-linear oscillators subject to external noise near a Hopf bifurcation instability and spontaneous structure formation in non-equilibrium systems, e.g., in cyclically competing populations or oscillatory chemical reactions. 
We employ a finite-difference method to numerically solve the noisy complex Ginzburg--Landau equation on a two-dimensional domain with the goal to investigate its non-equilibrium dynamics when the system is quenched into the ``defocusing spiral quadrant''. 
We observe slow coarsening dynamics as oppositely charged topological defects annihilate each other, and characterize the ensuing aging scaling behavior. 
We conclude that the physical aging features in this system are governed by non-universal aging scaling exponents. 
We also investigate systems with control parameters residing in the ``focusing quadrant'', and identify slow aging kinetics in that regime as well.
We provide heuristic criteria for the existence of slow coarsening dynamics and physical aging behavior in the complex Ginzburg--Landau equation.}

\begin{document}

\maketitle

\section{Introduction}
Macroscopic systems can be driven far away from thermal equilibrium states through a rapid change of external control parameters. 
If strong fluctuation effects are subsequently generated and govern the resulting relaxation processes, such massively perturbed systems may not be capable to rapidly return to a stationary state.
This is true especially in systems whose control parameters are quenched near a critical point where the typical relaxation time scales diverge \cite{Janssen89,Calabrese05,Tauber14}, as well as in a variety of disordered or dynamically frustrated systems that also display very slow relaxation kinetics, often characterized by scale-free power law decay \cite{Henkel11}. 
As a consequence, a significantly extended time window emerges wherein the systems retains memory to its initial state, and hence time translation invariance remains broken.
In this temporal regime one can observe typical ``aging'' phenomena, which describe the change of system properties over time with or without an applied external force, and provide a significant means to characterize complex collective behavior in a transient time regime long before (if ever) an asymptotic steady state has been reached \cite{Henkel11}. 
If in addition the slow relaxation processes are governed by a single algebraically growing (e.g., coarsening) length scale $L(t) \sim t^{1/z}$, dynamical scaling ensues in the aging regime. 
Thus, Ref.~\cite{Henkel11} defines physical aging phenomena in interacting many-body systems through three criteria: the presence of slow relaxation processes, e.g., algebraic rather than exponential relaxation; the breaking of time translation invariance; and the emergence of dynamical scaling. 
We also note the remarkable feature that physical aging is thermo-reversible even though time translation invariance does not hold.

Two-time quantities are often employed to characterize physical aging phenomena, such as the two-time auto-correlation function
\begin{equation}
    C(t,s) = \langle \phi(\bm{x},t) \, \phi(\bm{x},s) \rangle 
    - \langle \phi(\bm{x},t) \rangle \, \langle \phi(\bm{x},s) \rangle ,
    \label{eq:two_time_auto}
\end{equation}
where $\phi(\bm{x},t)$ denotes a suitably chosen time-dependent local order parameter, $t$ represents the ``observation'' time, and $s < t$ is called ``waiting'' time. 
Dynamical aging scaling may ensue in the limit $t \gg s \gg t_0$, where $t_0$ indicates any microscopic time scale; the auto-correlation function $C(t,s)$ then follows a ``simple aging'' scaling form if it satisfies
\begin{equation}
    C(t,s)=s^{-b}f _C(t/s) , 
    \label{eq:auto_scaling}
\end{equation}
where $b$ is termed aging scaling exponent, and $f_C(y)$ represents a scaling function which only depends on the time ratio $y = t/s$. In certain cases, there can be extensions to logarithmic rather than pure power law scaling \cite{Rutenberg95}.

Aging phenomena have been established to exist in a broad variety of physical systems \cite{Henkel11}, which range, e.g., from simple Ising ferromagnetics \cite{Zheng98, Henkel01, Pleimling04}, isotropic antiferromagnets \cite{Nandi19}, disordered magnets \cite{Henkel06}, spin glasses \cite{Henkel05}, disordered electronic Coulomb glass systems \cite{Shimer10,Shimer14,Assi16}, magnetic flux lines in type-II superconductors \cite{Nicodemi01, Bustingorry06, Bustingorry07,Pleimling11,Dobramysl13,Assi15,Pleimling15,Chaturvedi16}, disordered semiconductors \cite{Grempel04, Kolton05}, and skyrmion topological defects \cite{Brown18,Brown19} to driven lattice gases \cite{Daquila11,Daquila12}, population dynamics models \cite{Chen16}, and driven-dissipative Bose-Einstein condensation \cite{Liu16}. 
We remark that most of the above investigations utilized Monte Carlo or molecular dynamics computer simulations. 
In this present work, in contrast we provide a numerical exploration of aging scaling phenomena in the framework of a continuum non-linear partial differential equation, namely the intensely studied complex Ginzburg--Landau equation (CGL) \cite{Aranson02}. 
It provides a natural description of the kernel of many out-of-equilibrium physical systems that can be characterized by an amplitude (or ``envelope'' or ``modulational'') equation, yet includes only the simplest cubic non-linear term.
Nevertheless, the CGL describes an amazingly rich array of physical phenomena that include the spreading of non-linear waves, spontaneous pattern formation \cite{Cross93,Cross09}, and continuous dynamical phase transitions far away from thermal equilibrium.

\section{Model description}
Through straightforward rescaling, one may write the CGL in the following reduced form \cite{Aranson02} that is usually employed in numerical studies: 
\begin{eqnarray}
    \frac{\partial A(\bm{x},t)}{\partial t} &=& A(\bm{x},t) + (1+i\alpha) \nabla^2 A(\bm{x},t) \nonumber \\
    &&- (1+i\beta) |A(\bm{x},t)|^2 A(\bm{x},t) .
    \label{eq:cgle}
\end{eqnarray}
Here, $A(\bm{x},t)$ is a complex field, while $\alpha$ and $\beta$ are two real parameters that characterize the linear and non-linear dispersion, respectively. 
Eq.~(\ref{eq:cgle}) is an amplitude equation which yields a Hopf bifurcation in a spatially homogeneous setting.
In the spatially extended case, especially in two dimensions, it displays remarkably rich out-of-equilibrium behavior, including the presence of spatio-temporal chaos \cite{Coullet89a}, the nucleation of spiral wave structures \cite{Huber92}, and dynamic freezing \cite{Das12}. 
Therefore, it is reasonable to expect the emergence of aging phenomena for certain parameter ranges in this model system. Indeed, Chat\'e and Manneville pointed out that it should be promising to investigate aging in the frozen state of the two-dimensional CGL \cite{Chate96}; however, there are actually various problems in the observation of aging behavior in most of the states in this system, which we shall next discuss in detail.

First, we point out that fluctuating turbulent dynamical states in the two-dimensional CGL, irrespective of the presence or absence of topological defects, are not amenable to aging phenomena, as they are strongly disordering, and governed by fast relaxation processes. 
Specifically in two spatial dimensions, topological defects are identified by the zeros of the complex field $A(\bm{x},t)$. 
At these points, there exists a singularity of the phase $\theta(\bm{x},t) = {\rm arg} A(\bm{x},t)$, since it cannot be unambiguously defined for vanishing amplitude $|A(\bm{x},t)| = 0$. 
The integer topological charge $n = \oint_C d\theta / 2 \pi$, where $C$ indicates a closed contour around the zero of $A(\bm{x},t)$, quantifies these singularities. 
Only singly charged defects with $n = \pm 1$ are topologically stable. 
In the CGL with complex coefficients, these topological defects may emit persistent, well-established spiral waves, which ultimately fill the whole space and render the system ``frozen", i.e., terminate further temporal evolution. 
It turns out, as we have numerically checked in various instances, that the kinetic relaxation processes toward these frozen states also proceed too fast for aging phenomena to be observable.

Yet physical aging features were in fact previously observed in the two-dimensional classical XY model \cite{Abriet04}.
Indeed, the corresponding ``real'' Ginzburg--Landau (RGL) equation, which describes the ``model E'' kinetics  for a non-conserved two-component order parameter field \cite{Hohenberg77}, represents a special case of Eq.~(\ref{eq:cgle}) for $\alpha = 0 = \beta$.
It describes near-equilibrium relaxational kinetics for which oscillatory components are precluded.
The dynamics of the associated topological defects, namely vortices, in the RGL are easily understood in analogy with point-charge kinetics: 
Two oppositely charged topological defects will accelerate towards each other, and eventually this attraction will lead to the mutual annihilation of both defects. 
Similar dynamics should initially be expected for the CGL case as well. 
However, in the frozen state, the final surviving topological defects cannot actually react with each other because they remain  separated through stable shock structures between different well-established spiral domains. 
Information is then screened by these shocks and hence no interaction exists between defects in different domains. 
Therefore, to facilitate further slow relaxation kinetics, one should try to destroy these dynamically self-generated shock fronts. 
To this end, Das proposed to introduce disorder in the control parametres $\alpha$ or $\beta$ to effectively unlock the freezing-in \cite{Das12}.
Yet naturally this strategy will result in more complicated behavior for the characteristic growing length scale of the CGL system: 
Its time evolution will divide into two significantly distinct regimes, the first one corresponding to the transient freezing stage, whereas the second ``unlocking'' time window appears consistent with exponentially fast relaxation. 
Consequently, this strategy actually fails to open a slow aging time window. 

Interestingly, in a much earlier study from the same group, a power law growth of the characteristic length was observed for $\alpha = 0$ and $\beta$ set to either $0.25$ or $0.5$ \cite{Puri01}. 
The emergence of aging phenomena should thus be expected at these parameter values, for which according to Ref.~\cite{Puri01} the average length scale $L(t)$ of a single spiral structure becomes comparable to the wavelength of the associated asymptotic plane wave solutions. 
The spiral topological defects may then be expected to behave similarly to the vortices seen in the RGL or XY model. 
This condition is indeed satisfied either when the ``real'' Ginzburg-Landau limit is approached in the two parameter plane quadrants where $\alpha \beta < 0$ and the spirals are focusing, or when the system is quenched into the ``defocusing quadrant'' where $\alpha \beta > 0$. 
In their study of spiral domain patterns and shocks \cite{Bohr97}, Bohr et al. also suggested that if the length scale of a single spiral structure is not much larger than the characteristic wavelength, the phase matching constraint may be broken and no well-established shock lines formed, allowing for continuing slow mutal annihilation kinetics of oppositely  charged topological defects.

In the following, we study the aging behavior of the two-dimensional complex Ginzburg--Landau equation for both focusing and defocusing situations, which are distinguished by the winding orientation of the ensuing spiral topological defects. 
We first extract the dynamic exponent $z$, which describes the algebraic coarsening of the characteristic length scale $L(t) \sim t^{1/z}$, and subsequently determine the corresponding aging scaling exponents $b$ for the two-time auto-correlation function through appropriate data collapse for different waiting times $s$ for all the parameter choices we tested. 
We also try to calculate the auto-correlation exponent $\lambda_C$ that captures the long-time decay of the scaling function $f_c(y) \sim y^{-\lambda_C/z}$ for the parameter choices leading to (de-)focusing spirals \cite{Huse89}, utilizing $C(t,s=0)$ to extract the exponent $\lambda_C$.
Neverthelesss, as our systems can be tuned to approach the XY model limit in this case, this exponent will provide an opportunity for further comparison of these two models.

\section{Numerical scheme}
We employ a standard Euler discretization method to numerically integrate Eq.~(\ref{eq:cgle}), with a forward finite differential in time and a central differential for the Laplacian $\nabla^2$ in space. 
We exclusively consider systems on a two-dimensional square lattice with periodic boundary condition here, with system size $512 \times 512$, and discretization grid sizes $\Delta x=1$, $\Delta t=0.002$ in space and time, respectively, unless otherwise mentioned. 
We have also tested other system sizes and finer grids in time with $\Delta t=0.001$ case, which all provided results consistent with those reported in the following. 
We always employed fully randomly initialized configurations to ensure that a sufficient number of topological defects were present at the beginning of our numerical simulations. 
We also added a small Gaussian, white, and Markovian noise term $\eta(\bm{x},t)$ into Eq.~(\ref{eq:cgle}), which hence becomes a stochastic Langevin partial differential equation with weak additive noise $\eta(\bm{x},t)$, which is then fully characterized by the correlators
\begin{eqnarray}
    \langle \eta(\bm{x},t) \rangle &=& \langle \eta^*(\bm{x},t) \rangle = 0 \nonumber \\
    \langle \eta^*(\bm{x},t) \eta(\bm{x'},t') \rangle &=& \gamma \delta(\bm{x-x'}) \delta(t-t') \nonumber \\
    \langle \eta(\bm{x},t) \eta(\bm{x'},t') \rangle &=& \langle \eta^*(\bm{x},t) \eta^*(\bm{x'},t') \rangle = 0 ,
    \label{eq:add_noise}
\end{eqnarray}
where we chose $\gamma = 4.0\times 10^{-4}$ and $\eta^*(\bm{x},t)$ indicates the complex conjugate of $\eta(\bm{x},t)$. 

A characteristic coarsening length scale of our system, growing with time $t$, is given by average size of spiral domains
\begin{equation}
    l(t) = L / \sqrt{N(t)} ,
    \label{eq:char_length}
\end{equation}
where $L = 512$ is the linear extent of our simulation domain and $N(t)$ denotes the total number of topological defects in the system at time $t$. 
In order to investigate physical aging phenomena, we also calculate the two-time auto-correlation function for the complex order parameter $A(\bm{x},t)$, for which we need to slightly modify Eq.~(\ref{eq:two_time_auto}), and numerically compute the modulus 
\begin{equation}
    C(t,s) = | \langle A(\bm{x},t) A^*(\bm{x},s) \rangle | .
    \label{eq:complex_two_time_auto}
\end{equation}

\begin{figure}
\onefigure[width = 0.48 \textwidth]{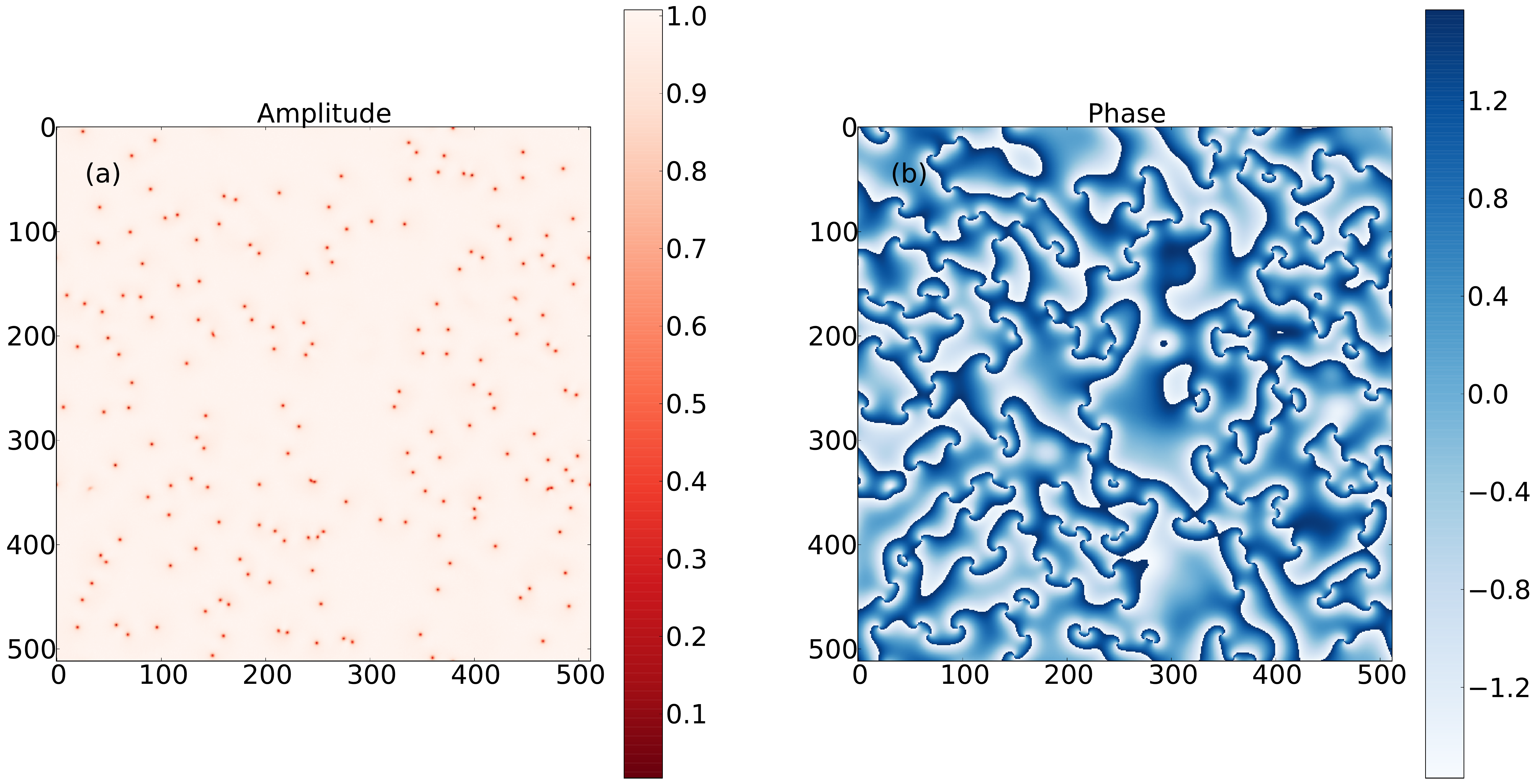}
\caption{An example of a CGL configuration obtained on a two-dimensional square lattice with post-quench parameters in the focusing spiral quadrant, and near the RGL limit with $\alpha = -0.05$, $\beta = 0.5$; at time $t = 500$: (a) shows the amplitude of the complex order parameter $A(\bm{x},t)$; (b) depicts its phase.}
\label{fig:ex_configuration}
\end{figure} 

\begin{figure*}[btp]
\onefigure[width = 0.9 \textwidth]{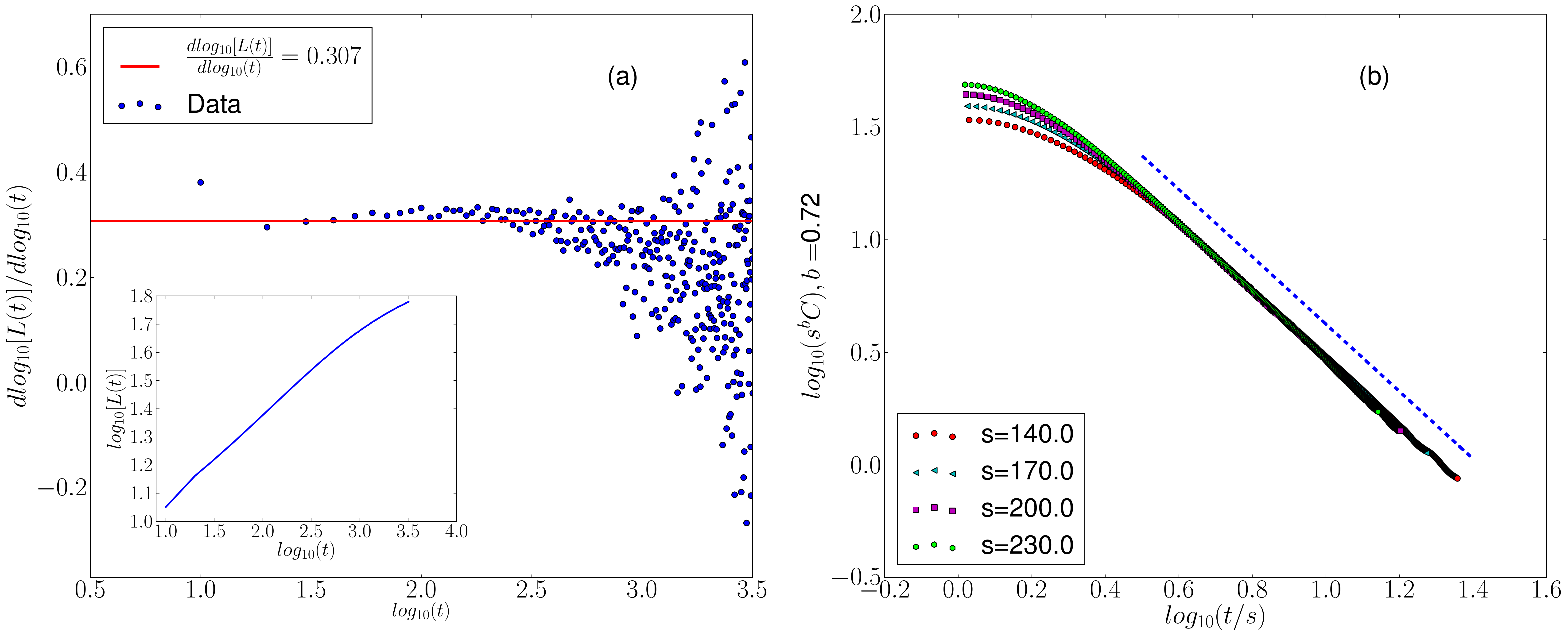}
\caption{(a) Derivative of the logarithm of the characteristic length $L(t)$ with respect to the logarithm of time $t$, i.e., effective local growth exponent $1/z(t)$ as function of $\log_{10} t$, for the two-dimensional CGL initiated with random configurations, set to control parameter values $\alpha = -0.05$ and $\beta = 0.5$, leading to focusing spiral structures.
The inset displays $log_{10}[L(t)]$ versus $log_{10}(t)$.
(b) Double-logarithmic plot of the scaled two-time auto-correlation function $s^b C(t,s)$ plotted as function of the time ratio $t/s$ for various waiting time $s$. 
Data collapse is achieved with aging scaling exponent $b = 0.72$. 
The dashed blue line with slope $- 1.56$ is approximately parallel to the collapsed curve, which indicates an auto-correlation / dynamic exponent ratio $\lambda_C / z \approx 1.56$. 
The data were obtained from averaging over $1000$ independent numerical simulation runs.}
\label{fig:ex_aging}
\end{figure*}

\section{Results}
In the following, we consider both the focusing spiral wave case when our CGL system is quenched near the RGL limit, as well as final parameter values residing in the defocusing quadrant. 
Fig.~\ref{fig:ex_configuration} shows an example of a numerically obtained CGL configuration for a system quenched to parameter values leading to focusing spirals. 
As evident in the right (blue-white) panel displaying the phase of the complex order parameter, there are small spiral structures present in the system which are however not well-established owing to the fact that their wavelength is comparable with their average size; hence there are no significant shock structures visible in the amplitude plot (left panel, red-white one). 
Note that the order parameter amplitude $|A(\bm{x},t)| \approx 1$ in most of the simulation domain, except at the (dark red) points that indicate the spatial locations of topological defects cores. 
There is no spontaneous creation of defect pairs in this case. 
Topological defects just attempt to reach their nearest neighbor with opposite charge and subsequently annihilate with it.
When the distance between two oppositely charged defects is sufficiently close, rather than directly approaching each other on a straight line like vortices in the XY model, they tend to drift tangentially, before gradually merging, and eventually annihilating. Some shock residuals are noticeable in the amplitude plot as the brightest regions indicating $|A(\bm{x},t)| > 1$. 
These nascent shocks become prominent upon departing from the RGL limit, and ultimately partition the CGL system into clearly separated and ``frozen'' spiral domains, terminating further large-scale dynamics. 
Defocusing spiral defects will behave similarly to the focusing case, except for the spirals rotating differently.

In Fig.~\ref{fig:ex_aging}(a), we show our data for the growth of the CGL characteristic length $L(t)$ with time for the control parameter pair $(\alpha,\beta) = (-0.05,0.5)$.
The effective local growth exponent $1/z(t)$, obtained from the logarithmic derivative of $L(t)$, remains stationary for a substantial time interval, and we can extract the inverse dynamic exponent $1/z \approx 0.307$; for large times $t > 1,000$ we naturally observe strong statistical fluctuations about this value, as finite-size effects become enhanced late in the time evolution.
The dynamic scaling exponent for the CGL appears clearly larger than its counterpart $z = 2$ for the RGL or XY model below the critical temperature, following a quench from a fully disordered initial state \cite{Henkel01}.
We note that in two dimensions, the latter also acquires a logarithmic correction in the functional relationship between characteristic length and the time, $L(t) \sim (t / \ln t)^{1/2}$ \cite{Rutenberg95, Jelic11}; see also Ref.~\cite{Comaron18} for an extension to non-equilibrium critical aging in a driven-dissipative quantum system.
However, if we attempt to apply a similar logarithmic correction to our CGL data, the resulting scaling behavior for $L(t)$ becomes slightly worse than with pure power law fits. 

As demonstrated in Fig.~\ref{fig:ex_aging}(b), we can also determine the aging scaling exponent $b \approx 0.8$ and the asymptotic decay exponent $\lambda_C / z \approx 1.56$ from the measured two-time auto-correlation function. 
Again compared with the two-dimensional XY model, for which $b \approx 0.03$ and $\lambda_C / z \approx 1.05$ were found at $T = 0.3 \, T_c $ \cite{Abriet04}, both our measured CGL aging exponent $b$ and auto-correlation exponent $\lambda_C / z$ values are markedly larger. 
These observations indicate growing length and non-equilibrium aging scaling behavior for the CGL systems with non-vanishing control parameter pair $(\alpha,\beta)$ that is clearly distinct from the classical XY model or RGL. 
We have explored the CGL aging scaling features more extensively through varying the control parameters $\alpha$ and $\beta$, averaging over $1,000$ independent realizations for each parameter pair. 
Table~\ref{table:1} lists our pertinent results for five different parameter sets. 
\begin{table}
\caption{Measured values of the dynamic, aging scaling, and auto-correlation exponents for the focusing spiral case.}
\label{table:1}
\begin{center}
\begin{tabular}{c|c|c|c}
\hline
$(\alpha,\beta)$ & $1/z$ & $b$ & $\lambda_C/z$  \\
\hline
\hline 
$(-0.05,0.5)$ & $0.307(3)$ & $0.72(6)$ & $1.56(11)$\\
\hline
$(-0.04,0.6)$ & $0.293(4)$ & $0.38(5)$ & $1.31(9)$\\
\hline
$(-0.04,0.5)$ & $0.312(3)$ & $0.80(5)$ & $1.63(13)$\\
\hline
$(-0.06,0.5)$ & $0.298(4)$ & $0.68(6)$ & $1.49(17)$\\
\hline
\end{tabular}
\end{center}
\end{table}

\begin{figure*}[btp]
\onefigure[width = 0.9 \textwidth]{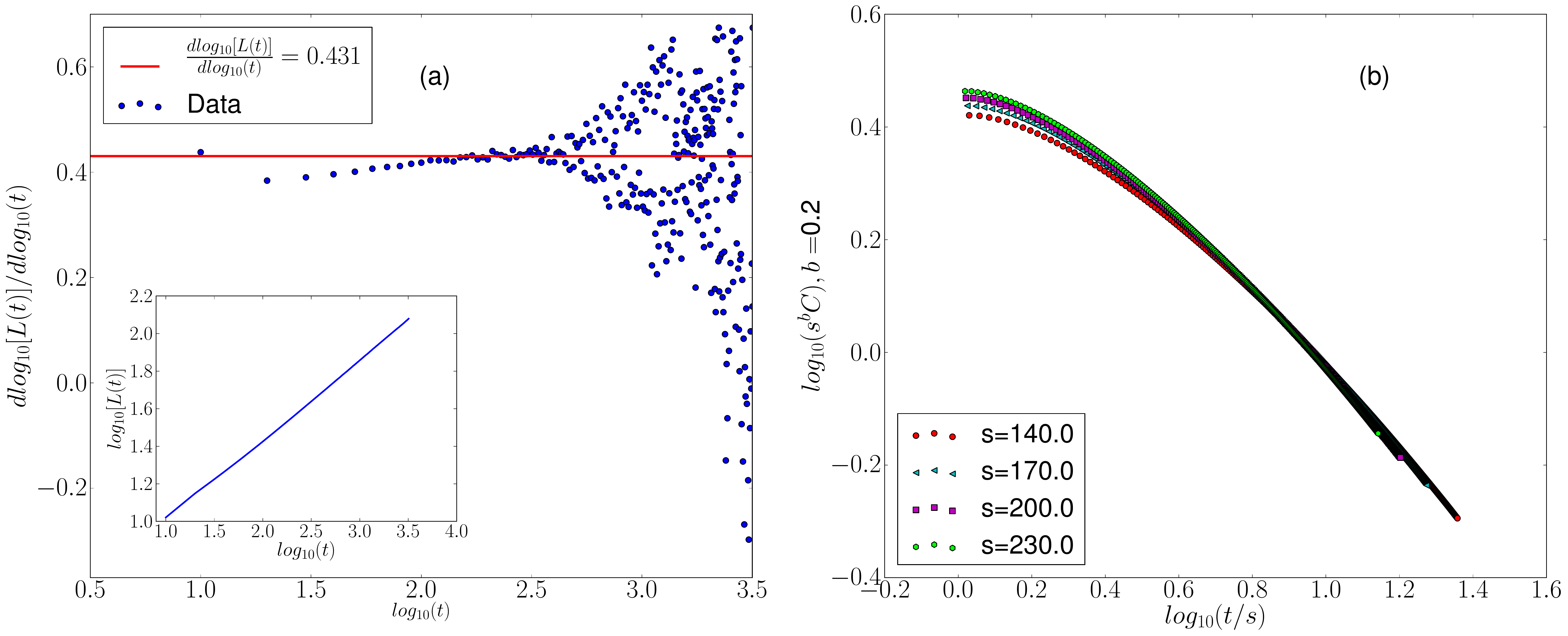}
\caption{(a) Effective local growth exponent $1/z(t)$ for the characteristic length $L(t)$ as function of $\log_{10} t$ for the two-dimensional CGL with post-quench control parameters $\alpha=1.176$ and $\beta=0.7$ in the defocusing quadrant.
The inset displays $log_{10}[L(t)]$ versus $log_{10}(t)$.
(b) Scaled two-time auto-correlation function vs. $t/s$ for various waiting time $s$, with aging scaling exponent $b=0.2$. 
The data resulted from averaging over $8000$ independent runs.}
\label{fig:ex_aging_defocusing}
\end{figure*}
In our investigation of potential aging scaling in the CGL system, we have focused on small values of $\alpha$ case as this parameter is directly related to a characteristic inverse spatial scale. 
Our extracted values of the aging exponents $b$ suggest non-universal aging scaling behavior of CGL systems in the control parameter sector leading to focusing spirals. 
This is reasonable since the non-vanishing linear term in Eq.~(\ref{eq:cgle}) ensures that the system resides below the critical point. 
We observe enhanced variance for the measured exponents when the parameter $\beta$ is changed at fixed small values of $\alpha$, while we only see minor changes if we vary $\alpha$ for fixed $\beta$. 
The origin of this difference is that the control parameter $\beta$ contributes much more than $\alpha$ to effectively cause deviations from the RGL limit $\alpha = 0 = \beta$. 
Furthermore, the system with parameter pair $(-0.04,0.6)$ becomes almost frozen at the very end of the simulation runs.
Correspondingly, we find the scaled two-time auto-correlation functions curves to be dispersed for this parameter choice. 

For small $\alpha$ values in the focusing spiral quadrant, one explicitly approaches the RGL limit; alternatively, we may also eliminate significant shock structures and thus generate interactive spiral dynamics through quenching the CGL into the defocusing quadrant, where $\alpha \beta > 0$. 
In this case, it is not necessary to tune the control parameters near the origin in the $(\alpha,\beta)$ parameter plane. 
Figure~\ref{fig:ex_aging_defocusing}(a) displays an example for the ensuing characteristic length scale growth exponent for
$(\alpha,\beta) = (1.176,0.7)$.
The stationary value $1/z \approx 0.431$ is clearly different from the focusing spiral quadrant, implying distinct relaxational dynamics for the topological defects in both regimes. 
The associated scaled two-time auto-correlation functions are shown in Fig.~\ref{fig:ex_aging_defocusing}(b). 
The curves display noticeable bending even in the regime where the data allow for aging scaling collapse, implying that the system might not have reached the asymptotic simple power law decay $f_c(y) \sim y^{-\lambda_C / z}$ for the scaling function; hence we report tentative values for the auto-correlation exponent $\lambda_C$ using $C(t,s=0)$ here.

\begin{table}
\caption{Measured values of the dynamic and aging scaling exponents for the defocusing spiral case.}
\label{table:2}
\begin{center}
\begin{tabular}{c|c|c|c}
\hline
$(\alpha,\beta)$ & $1/z$ & $b$ & $\lambda_C/z$\\
\hline
\hline 
$(1.176,0.7)$ & $0.410(3)$ & $0.20(5)$ & $0.71(3)$\\
\hline
$(1.176,0.65)$ & $0.406(3)$ & $0.24(3)$ & $0.78(6)$\\
\hline
$(1.176,0.55)$ & $0.394(3)$ & $N/A$ & $ 1.15(10)$\\
\hline
$(1.111,0.6)$ & $0.416(3)$ & $0.23(5)$ & $0.78(6)$\\
\hline
$(1.429,0.55)$ & $0.388(3)$ & $N/A$ & $ 1.92(34)$\\
\hline
$(1.429,0.6)$ & $0.394(3)$ & $N/A$ & $ 1.41(17)$\\
\hline
$(1.250,0.55)$ & $0.392(3)$ & $N/A$ & $1.20(15)$\\
\hline
$(1.250,0.6)$ & $0.399(3)$ & $N/A$ & $0.97(9)$\\
\hline
\end{tabular}
\end{center}
\end{table}
In Table~\ref{table:2} we collect the resulting scaling exponent values for eight different sets of parameters. 
Except for the $(1.176,0.7)$ pair, all other results were obtained through averaging over $1000$ independent realizations. 
The measured dynamical exponents $z$ are generally smaller and depend rather weakly on the control parameters in the explored range, as compared with the focusing spiral cases. 
Considering possible aging scaling for the two-time auto-correlation function, we find that the possible ranges of the aging collapse exponents $b$ do have overlaps. 
However, due to the fact that aging scaling cannot reliably be demonstrated for five of these parameter pairs (indicated by the ``N/A'' entries in Table~\ref{table:2}), we wish to more carefully assess possible conclusions about the universality of the exponent $b$.
We also observe large variations in the extracted values of $\lambda_C / z$ for these five parameter pairs, with similar overlap ranges for the intervals defined by our measurement errors as for the three parameter sets for which we could reliably extract the aging scaling exponent $b$.

To this end, we want to first address the question under which circumstances we would expect dynamical scaling behavior in the CGL, and generally spirals may behave akin to vortices even when the system is quenched far from the RGL limit $\alpha = 0 = \beta$. 
In fact, one may reduce Eq.~(\ref{eq:cgle}) to the RGL limit in the special case $\alpha = \beta \ne 0$ through a transformation into a ``rotating'' frame $A \to A e^{i \beta t}$, if one is interested in an approximate analytic solution for an  isolated spiral \cite{Aranson02}. 
Upon neglecting higher-order corrections, the resulting solution is identical with one for the RGL solution except for a wavenumber shift. 
Consequently there exists a whole branch of vortex-like behavior in the defocusing quadrant in the $\alpha \to \beta$ limit, which is not the case when $\alpha \beta < 0$. 
Hence we conjecture that CGL systems which are quenched into the defocusing spiral quadrant far from the origin in the $(\alpha,\beta)$ parameter space could still display proper dynamical aging scaling, provided their control parameter pair is set near the line $\alpha = \beta$.
Indeed, our numerical results listed in Table~\ref{table:2} are suggestive as well that aging scaling is restored when the difference of control parameters approaches $\alpha - \beta \to 0$. 
Since the asymptotic spiral wavenumber vanishes in this limit, the effect of shock structures should be expected to be weak.
However, as long as $\alpha = \beta \ne 0$, there remains an obvious rotation of the equiphase lines, and the interaction between defects, either repulsive or attractive, will be oblique \cite{Aranson02}. 
Thus two defects will never move towards or away from each other along the line connecting them, like the vortices in the RGL case. 
Instead they will either spiral around each other if they carry the same topological charge, or move along the tangential direction, approaching gradually and finally annihilating with each other, if they are oppositely charged.

\begin{table}
\caption{Measured values of the dynamic and aging scaling exponents for the defocusing spiral case, near the limit $\alpha =\beta$.}
\label{table:3}
\begin{center}
\begin{tabular}{c|c|c|c}
\hline
$(\alpha,\beta)$ & $1/z$ & $b$ & $\lambda_C/z$\\
\hline
\hline 
$(1.2,1.2)$ & $0.441(2)$ & $0.04(4)$ & $0.59(3)$\\
\hline
$(1.0,1.0)$ & $0.426(3)$ & $0.04(3)$ & $0.59(3)$\\
\hline
$(1.2,1.25)$ & $0.444(2)$ & $0.05(4)$ & $0.60(4)$\\
\hline
$(1.2,1.15)$ & $0.440(3)$ & $0.04(3)$ & $0.60(3)$\\
\hline
\end{tabular}
\end{center}
\end{table}
We have further tested this hypothesis by explicitly quenching the CGS control parameters near the $\alpha = \beta$ limit; our results for four such parameter pairs are listed in Table~\ref{table:3}. 
First we observe that the aging scaling collapse exponent $b \approx 0.03$ from a previous study of the two-dimensional XY model below the critical temperature \cite{Abriet04} falls into all the extracted likely ranges for $b$ in the complex CGL reported in this table. 
This confirms the conclusion that the CGL non-equilibrium relaxation kinetics will (partly) restore XY model dynamical scaling features behaviors when $\alpha \to \beta \ne 0$. 
Second, the inverse dynamical scaling exponent is also seen to be closer to the XY limit $1/z = 0.5$ (with logarithmic corrections in two dimensions) than the values reported in Tables~\ref{table:1} and \ref{table:2}. 
The measured small deviation might either stem from truly distinct dynamical scaling behavior in these CGL systems as compared to the RGL, or could perhaps be attributed to sizeable logarithmic corrections.
Yet attempts to include logarithmic terms in the aging analysis again did not improve the scaling properties.
Further detailed studies extending to longer simulation times, and hence requiring larger system sizes, may be needed to confirm either scenario. 
Finally, although the aging scaling exponents in Table~\ref{table:3} share similar ranges, they are clearly different from those reported in Table~\ref{table:2}, which suggest non-universal aging behavior of the spirals in the defocusing quadrant as well.
Similar observations and conclusions pertain to the associated auto-correlation decay exponents.
Yet the double-logarithmic plots of the scaled auto-correlation functions still show marked curvature, so this feature persists even in the limit $\alpha \to \beta$ in our observation time window. 
We tentatively propose that this slower approach to the long-time asymptotic may be caused by the defocusing rotation of the equiphase lines of the resulting spiral structures, which will decelerate their dynamics.

\section{Conclusion}
In summary, we have investigated the emergent aging phenomena and ensuing dynamic scaling in certain parameter regions for the two-dimensional complex Ginzburg--Landau equation, both in the focusing and defocusing spiral quadrants. 
To avoid shock structures which will prevent the interaction between defects and cause the system to become spatially frozen, we have argued that one should quench the control parameters near the RGL limit $\alpha = 0 = \beta$ in the focusing spiral case, or alternatively be in the defocusing quadrant. 
In the time windows we have considered here, both the dynamical scaling and aging exponents differ slightly from the corresponding behavior of coarsening vortices in the two-dimensional XY model, when the former system is quenched below the critical temperature in the focusing spiral regime, even when the XY model limit is approached. 
In fact, for certain control parameter pairs, we could not even observe any aging features, which provides evidence that these systems display non-universal non-equilibrium relaxation dynamics.
 
In the defocusing spiral region, one may approach another special situation, namely $\alpha = \beta$ while avoiding the regime $\alpha = 0 = \beta$; aging phenomena are then prevalent even if the systems are quenched far away from the origin in the parameter space, but with non-universal scaling exponents. 
We invariable detect a minute bending of the scaled two-time auto-correlation function curves in all our defocusing spiral case studies, independent if aging scaling emerges or not. 
We tentatively attribute this feature as to the defocusing rotation of the equi-phase line connecting different topological defects (or vortices). 
Thus, no simple aging scaling behavior appears to be feasible in this defocusing spiral quadrant, except near the line $\alpha = \beta$ in parameter space.
However, numerical data with considerably improved statistics, as well as much larger system sizes, may be required to unambiguously confirm this conclusion.
Since the non-equilibrium relaxation behavior of these two-dimensional CGL systems turned out quite non-universal, it would also be interesting and important to investigate the effects of different boundary conditions as well as other, correlated initial configurations, e.g., a uniform initial distribution with a small inhomogeneous perturbation.

\acknowledgments
The authors are indebted to Bart Brown, Harshwardhan Chaturvedi, and Michel Pleimling for helpful discussions, and to Jacob Carroll for a careful reading of the manuscript draft.
This research is supported by the U.S. Department of Energy, Office of Basic Energy Sciences, Division of Materials Science and Engineering under Award DE-SC0002308.






\begin{thebibliography}{10}
\expandafter\ifx\csname url\endcsname\relax\def\url#1{\texttt{#1}}\fi

\bibitem{Janssen89}
\Name{Janssen H. K., Schaub B. \and Schmittmann B.} \REVIEW{Z. Phys. B Condens. Matter}{73}{1989}{539}.

\bibitem{Calabrese05}
\Name{Calabrese P. \and Gambassi A.} \REVIEW{J. Phys. A: Math. Gen.}{38}{2005}{R133}.

\bibitem{Tauber14}
\Name{T{\"a}uber U.~C.} \Book{Critical dynamics: a field theory approach to
  equilibrium and non-equilibrium scaling behavior} (Cambridge: Cambridge
  University Press) 2014.

\bibitem{Henkel11}
\Name{Henkel M. \and Pleimling M.} \Book{Non-Equilibrium Phase Transitions:
  Volume 2: Ageing and Dynamical Scaling Far from Equilibrium} (Springer
  Science \& Business Media) 2011.

\bibitem{Rutenberg95}
\Name{Rutenberg A.~D. \and Bray A.~J.} \REVIEW{Phys. Rev. E}{51}{1995}{5499}.

\bibitem{Zheng98}
\Name{Zheng B.} \REVIEW{Int. J. Mod. Phys. B}{12}{1998}{1419}.

\bibitem{Henkel01}
\Name{Henkel M., Pleimling M., Godr{\`e}che C. \and Luck J.-M.} \REVIEW{Phys. Rev. Lett.}{87}{2001}{265701}.

\bibitem{Pleimling04}
\Name{Pleimling M.} \REVIEW{Phys. Rev. B}{70}{2004}{104401}.

\bibitem{Nandi19}
\Name{Nandi R. \and T{\"a}uber U.~C.} \REVIEW{Phys. Rev. B}{99}{2019}{064417}.

\bibitem{Henkel06}
\Name{Henkel M. \and Pleimling M.} \REVIEW{Europhys. Lett.}{76}{2006}{561}.

\bibitem{Henkel05}
\Name{Henkel M. \and Pleimling M.} \REVIEW{Europhys. Lett.}{69}{2005}{524}.

\bibitem{Shimer10}
\Name{Shimer M.~T., T{\"a}uber U.~C. \and Pleimling M.} \REVIEW{Europhys. Lett.}{91}{2010}{67005}.

\bibitem{Shimer14}
\Name{Shimer M.~T., T{\"a}uber U.~C. \and Pleimling M.} \REVIEW{Phys. Rev. E}{90}{2014}{032111}.

\bibitem{Assi16}
\Name{Assi H., Chaturvedi H., Pleimling M. \and T{\"a}uber U.~C.} \REVIEW{Eur. Phys. J. B}{89}{2016}{252}.

\bibitem{Nicodemi01}
\Name{Nicodemi M. \and Jensen H.~J.} \REVIEW{J. Phys. A: Math. Gen.}{34}{2001}{8425}.

\bibitem{Bustingorry06}
\Name{Bustingorry S., Cugliandolo L.~F. \and Dom{\'i}nguez D.} \REVIEW{Phys. Rev. Lett.}{96}{2006}{027001}.

\bibitem{Bustingorry07}
\Name{Bustingorry S., Cugliandolo L.~F. \and Dom{\'i}nguez D.} \REVIEW{Phys. Rev. B}{75}{2007}{024506}.

\bibitem{Pleimling11}
\Name{Pleimling M. \and T{\"a}uber U.~C.} \REVIEW{Phys. Rev. B}{84}{2011}{174509}.

\bibitem{Dobramysl13}
\Name{Dobramysl U., Assi H., Pleimling M. \and T{\"a}uber U.~C.} \REVIEW{Eur. Phys. J. B}{86}{2013}{228}.

\bibitem{Assi15}
\Name{Assi H., Chaturvedi H., Dobramysl U., Pleimling M. \and T{\"a}uber U.~C.}
  \REVIEW{Phys. Rev. E}{92}{2015}{052124}.

\bibitem{Pleimling15}
\Name{Pleimling M. \and T{\"a}uber U.~C.} \REVIEW{J. Stat. Mech.}{2015}{2015}{P09010}.

\bibitem{Chaturvedi16}
\Name{Chaturvedi H., Assi H., Dobramysl U., Pleimling M. \and T{\"a}uber U.~C.}
  \REVIEW{J. Stat. Mech.}{2016}{2016}{083301}.

\bibitem{Grempel04}
\Name{Grempel D. R.} \REVIEW{Europhys. Lett.}{66}{2004}{854}.

\bibitem{Kolton05}
\Name{Kolton A.~B., Grempel D. R. \and Dom{\'\i}nguez D.} \REVIEW{Phys. Rev. B}{71}{2005}{024206}.

\bibitem{Brown18}
\Name{Brown B.~L., T{\"a}uber U.~C. \and Pleimling M.} \REVIEW{Phys. Rev. B}{97}{2018}{020405(R)}.

\bibitem{Brown19}
\Name{Brown B.~L., T{\"a}uber U.~C. \and Pleimling M.} \REVIEW{Phys. Rev. B}{100}{2019}{024410}.

\bibitem{Daquila11}
\Name{Daquila G.~L. \and T{\"a}uber U.~C.} \REVIEW{Phys. Rev. E}{83}{2011}{051107}.

\bibitem{Daquila12}
\Name{Daquila G.~L. \and T{\"a}uber U.~C.} \REVIEW{Phys. Rev. Lett.}{108}{2012}{110602}.

\bibitem{Chen16}
\Name{Chen S. \and T{\"a}uber U.~C.} \REVIEW{Phys. Biol.}{13}{2016}{025005}.

\bibitem{Liu16}
\Name{Liu W. \and T{\"a}uber U.~C.} \REVIEW{J. Phys. A: Math. Theor.}{49}{2016}{434001}.

\bibitem{Aranson02}
\Name{Aranson I.~S. \and Kramer L.} \REVIEW{Rev. Mod. Phys.}{74}{2002}{99}.

\bibitem{Cross93}
\Name{Cross M.~C. \and Hohenberg P.~C.} \REVIEW{Rev. Mod. Phys.}{65}{1993}{851}.

\bibitem{Cross09}
\Name{Cross M. \and Greenside H.} \Book{Pattern formation and dynamics in
  nonequilibrium systems} (Cambridge: Cambridge University Press) 2009.

\bibitem{Coullet89a}
\Name{Coullet P., Gil L. \and Lega J.} \REVIEW{Phys. Rev. Lett.}{62}{1989}{1619}.

\bibitem{Huber92}
\Name{Huber G., Alstr{\o}m P. \and Bohr T.} \REVIEW{Phys. Rev. Lett.}{69}{1992}{2380}.

\bibitem{Das12}
\Name{Das S.~K.} \REVIEW{Europhys. Lett.}{97}{2012}{46006}.

\bibitem{Chate96}
\Name{Chat{\'e} H. \and Manneville P.} \REVIEW{Physica A}{224}{1996}{348}.

\bibitem{Abriet04}
\Name{Abriet S. \and Karevski D.} \REVIEW{Eur. Phys. J. B}{37}{2004}{47}.

\bibitem{Hohenberg77}
\Name{Hohenberg P.~C. \and Halperin B.~I.} \REVIEW{Rev. Mod. Phys.}{49}{1977}{435}.

\bibitem{Puri01}
\Name{Puri S., Das S.~K. \and Cross M. C.} \REVIEW{Phys. Rev. E}{64}{2001}{056140}.

\bibitem{Bohr97}
\Name{Bohr T., Huber G. \and Ott E.} \REVIEW{Physica D}{106}{1997}{95}.

\bibitem{Huse89}
\Name{Huse D.~A.} \REVIEW{Phys. Rev. B}{40}{1989}{304}.

\bibitem{Jelic11}
\Name{Jeli{\'c} A. \and Cugliandolo L.~F.} \REVIEW{J. Stat. Mech.}{2011}{2011}{P02032}.

\bibitem{Comaron18}
\Name{Comaron P., Dagvadorj G., Zamora A., Carusotto I., Proukakis N.~P. \and Szyma{\'n}ska M.~H.} 
\REVIEW{Phys. Rev. Lett.}{121}{2018}{095302}.

\end{thebibliography}

\end{document}